\documentclass[%
reprint,
 amsmath,amssymb,
 aps,
]{revtex4-2}

\usepackage{graphicx}
\usepackage{dcolumn}
\usepackage{bm}

\usepackage{booktabs}
\usepackage{xcolor}
\usepackage{float}

\newcommand {\fabsq}[1] {\left\vert #1 \right\vert^2}

\newcommand{\ket}[1]{\ensuremath{|#1\rangle}}
\newcommand{\bra}[1]{\langle#1|}

\newcommand{\braket}[2]{\langle#1|#2\rangle}

\newcommand{\braXket}[3]{\langle#1|#2|#3\rangle}

\newcommand{\cH}{{\cal{H}}}

\newcommand{\Eqref}[1]{Eq. \eqref{#1}}
\newcommand{\Hamil}{\mathcal{H}}

\newcommand\imageSizeA{0.325}
\newcommand\imageSizeB{0.48}
\newcommand\imageSizeC{0.49}

\newcommand\figVspace{-20}

\begin{document}

\preprint{APS/123-QED}

\title{Quantum control of a random transverse Ising spin system}

\author{C. Whitty}
 \email{c.whitty@ehu.eus}
\author{E. Ya. Sherman}
\affiliation{%
 Department of Physical Chemistry, University of the Basque Country UPV/EHU, 48940 Leioa, Spain}
 
\author{Xi Chen}
\author{Yue Ban}
\affiliation{Instituto de Ciencia de Materiales de Madrid (CSIC), Cantoblanco, E-28049 Madrid, Spain}
 
\date{\today}

\begin{abstract}
We consider subspace transfer within the time-dependent one-dimensional quantum transverse Ising model, with random nearest-neighbor interactions and a transverse field. We run numerical simulations using a variational approach and the numerical GRAPE (gradient-ascent pulse engineering) and dCRAB (dressed chopped random basis) quantum control algorithms.
\end{abstract}

\maketitle

Quantum many-body systems exhibit a plethora of phenomena that have no analogue in classical systems and offer unprecedented opportunities for designing and fabricating future quantum technologies, for example, quantum sensing, quantum computing and new materials \cite{montenegroReviewQuantumMetrology2024,fausewehQuantumManybodySimulations2024,goyalExploringQuantumMaterials2025,acinQuantumTechnologiesRoadmap2018a,mukherjeePromisesChallengesManybody2024}.

We consider time-dependent control of a quantum Ising model with random nearest-neighbor interactions and a control transverse field.
This model has been extensively studied, and we use it as a platform to investigate subspace-to-subspace transfer \cite{fisherRandomTransverseField1992a,florencioDynamicsRandomOnedimensional1999,fytasReviewRecentDevelopments2018, duttaQuantumPhaseTransitions2015a}.

There are many techniques to design control schemes for the transverse Ising model without random interactions, for example counter-diabatic driving, quantum quench protocols, and quantum annealing \cite{guery-odelinShortcutsAdiabaticityConcepts2019,mitraQuantumQuenchDynamics2018,rajakQuantumAnnealingOverview2022a, quantumannealing, bernaschiQuantumTransitionTwodimensional2024a}.
Quantum spin glasses and quantum annealing cover a large area of research, where control of quantum spin glasses can provide a platform for adiabatic computation \cite{quantumannealing, albashAdiabaticQuantumComputation2018a,jaumaExploringQuantumAnnealing2024}.
Dynamical invariants can be used to cross phase transitions in a transverse Ising model with small randomness in nearest-neighbor couplings, but this control breaks down in the case of strong randomness \cite{espinosInvariantbasedControlQuantum2023}.
Robust state preparation has been implemented for small noise coupling in spin chains, where the control problem is written implicitly using dynamical invariants allowing numerical optimization without requiring an analytic expression for the invariant \cite{stefanescuRobustImplicitQuantum2024,orozco-ruizQuantumControlQuantum2024a}.

\begin{figure}
\flushleft a)\\ \vspace{-10 pt}
\centering
\includegraphics[width=0.9\linewidth]{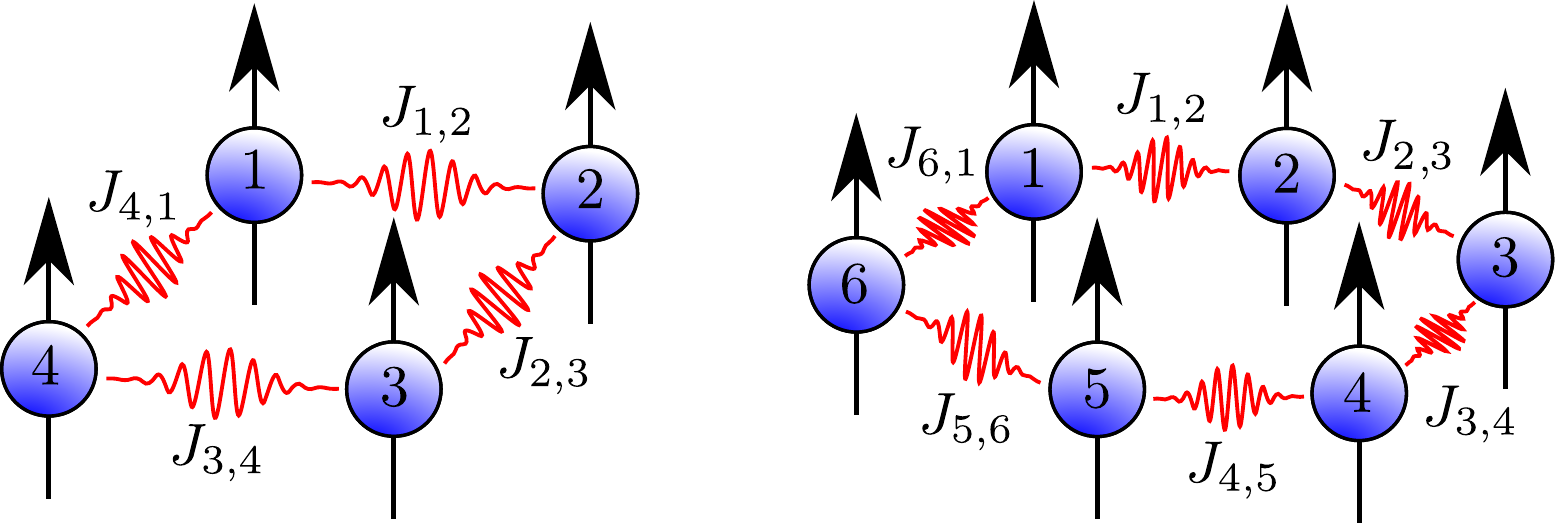}
\flushleft b)\\ \vspace{-10 pt}
\centering
\includegraphics[width=1\linewidth]{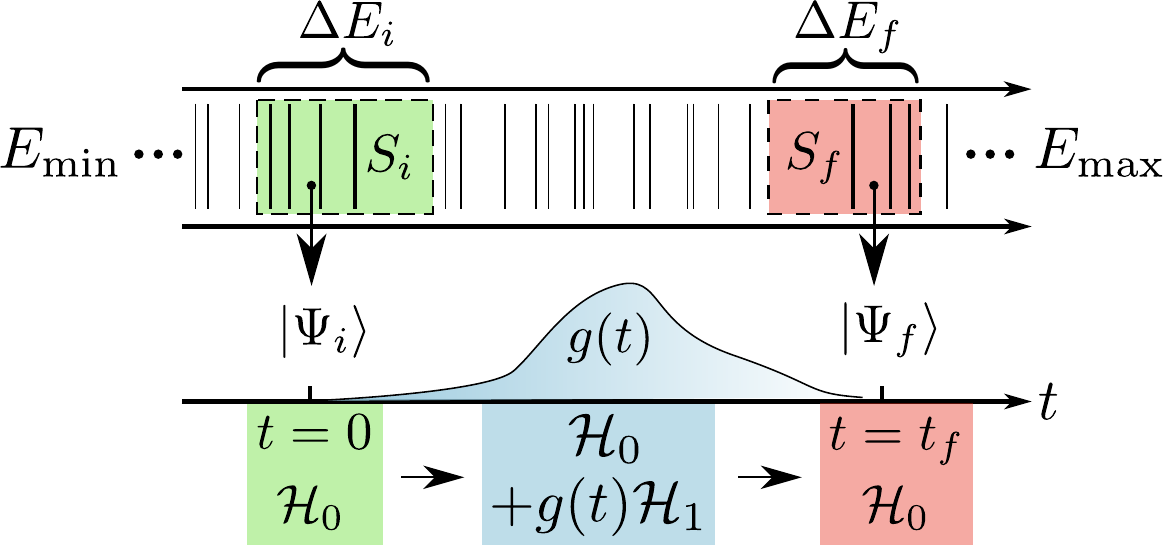}
\caption{a) Diagram of $N=4,6$ spins with nearest neighbor interactions $J_{i,i+1}$, and periodic boundary conditions.\\
b) Schematic depiction of the energy spectrum and control problem. The energy spectrum is represented by the top row, where $\ket{\Psi_i}\in S_i$ and $\ket{\Psi_f}\in S_f$. The vertical lines represent eigenenergies with $\Delta E_k$ being the difference between the largest and smallest eigenenergy in each $S_k$. The second row depicts the time evolution, with $\Hamil(0)=\Hamil_0=\Hamil(t_f)$, and during $0<t<t_f$, $\Hamil(t)=\Hamil_0+g(t)\Hamil_1$. A possible pulse shape is shown as $g(t)$.}
\label{fig:setup}
\end{figure}

Our goal is to maximize state transfer between two energy eigenstate subspaces of a transverse Ising model with random nearest-neighbor couplings, see Fig. \ref{fig:setup} for a schematic of the control problem.
We consider several techniques to design control pulses in this strongly random regime, starting with simple pulse shapes, and then implement GRAPE (gradient ascent pulse engineering) and CRAB (chopped random basis) optimal control algorithms \cite{khanejaOptimalControlCoupled2005,mullerOneDecadeQuantum2022, rossignoloQuOCSQuantumOptimal2023}.

The paper is laid out as follows; in Section II we describe the model, in Section III we show the effectiveness of the different control schemes, and then we discuss our findings in the Conclusion.

\section{\label{sec:model_hamiltonian}Model Hamiltonian}
Our goal is to control a random coupling transverse Ising system of spin $1/2$ particles, and Hamiltonian
\begin{align}
H(t) &= \Hamil_0 + g(t) \Hamil_1 + \beta \Hamil_2,
\end{align}
where
\begin{align}
&\Hamil_0 = -\sum_{i=1}^N J_{i,i+i} \sigma_i^z \sigma_{i+1}^z,
\nonumber \\
&\Hamil_1 = -\sum_{i=1}^N \sigma_i^x,
 \quad\Hamil_2 = \sigma_i^z,
\end{align}
and $J_{i,i+1}$ are random couplings between neighboring spins chosen from a uniform distribution, $g(t)$ is the real-valued time-dependent control of the transverse field, $N$ is the number of spins, $\beta$ is a small real-valued degeneracy-breaking parameter and $\sigma^\alpha$ are the usual Pauli operators.
We will use periodic boundary conditions, i.e. $\sigma^z_{N} \sigma^z_{N+1} = \sigma^z_{N} \sigma^z_{1}$.
Our task is to design a control scheme $g(t)$ that transfers an initial state $\ket{\Psi_i}\in S_i$ to a final state $\ket{\Psi_f}\in S_f$, in a given time $t_f$, and where $S_i,S_f$ are subspaces defined by energy bands $\Delta E_i,\Delta E_f$ respectively.
We set $g(0)=0$ and $g(t_f)=0$, so that $H(0)=H(t_f)$.

We consider a closed system, and solve the time-dependent Schr\"odinger equation
\begin{align}
i \hbar \frac{\partial}{\partial t} \ket{\Psi(t)} = H(t) \ket{\Psi(t)},
\end{align}
with $\ket{\Psi(0)}=\ket{\Psi_i}$, $\ket{\Psi(t_f)}=\ket{\Psi_f}$.
To measure the effectiveness of $g(t)$ we will use the standard state fidelity measure $F = \fabsq{\braXket{\Psi_f}{U\left( t_f,0 \right)}{\Psi_i}}$, where $U$ is the time evolution operator.
To measure the effectiveness of subspace transfer, we define a fidelity measure $F_S$ given by
\begin{align}
F_{S} = \fabsq{
\sum_{k=1}^K \sum_{j=1}^J\braket{\Psi_f}{\phi_k}\bra{\phi_k}
U\left( t_f,0 \right) \ket{\phi_j}\braket{\phi_j}{\Psi_i}
},
\end{align}
where $S_i = \text{Span} \{ \phi_k : k=1,\dots,K\}$ and $S_f = \text{Span} \{ \phi_j : j=1,\dots,J\}$.
For simplicity, we assume $S_i \cap S_f = \emptyset$ and we set the initial state $\ket{\Psi_i} $ and the target state $\ket{\Psi_f} $ to be
\begin{align}
\ket{\Psi_i} &= c_i\sum_{k=k_n}^{k_m} \ket{k},
&\ket{\Psi_f} = c_f\sum_{k=k_l}^{k_s} \ket{k},
\end{align}
where $\ket{k}$ are energy eigenstates of $H_0 +\beta H_2$ and $c_i,c_f$ are normalization constants, with no overlap between $\{k_n,k_n+1,\dots,k_m-1,k_m\}$ and $\{k_l,k_l+1,\dots,k_s-1,k_s\}$, .i.e $\braket{\Psi_i}{\Psi_f}=0$.

In this paper we use even values of $N$ and let $k_n=2^N \times 3/16$, $k_m=2^N \times 6/16$, $k_l=2^N \times 11/16$ and $k_s=2^N \times 14/16$, with $c_{i,f}=(3/16\times 2^N +1) \textsuperscript{$\wedge$} (-1/2)$.
In this setting $F=F_S$, so we use $F$ to denote the fidelity achieved with these initial and target states.
Choosing these initial and target states gives a reasonable comparison in terms of portions of the spectrum from which we want to achieve transfer, for different $N$.

We consider a uniform distribution for the nearest-neighbor couplings $J_{i,i+1}\in[-1,1]$, where the values have been chosen once and used again for comparison in different control schemes.

\subsection{Spectrum of $\Hamil_0 + \beta\Hamil_z$}
We consider the energy spectrum of the system with the pulse off i.e., $g(0)=0$, and taking a standard matrix representation of the Pauli matrices we calculate the eigenvalues.
With $\beta \ne 0$ and $\Hamil_z = \sum_{i=1}^N \sigma_i^z$, there exists $\begin{pmatrix} N-1 \\ N/2 \end{pmatrix}$ degenerate eigenvalues for $N$ even, and no degeneracies for $N$ odd.
We find that choosing $\beta \times \Hamil_z =0.001 \times \sigma_1^z$ removes all degeneracies for both even and odd $N$.

For a given set of $J_{i,i+1}$ and Hamiltonian $H_0 + \beta \Hamil_z$, the energy gap between states $\ket{\phi_k}$ and $\ket{\phi_{k+1}}$ can be found, with a symmetry in the spectrum gaps about the middle.
For all $N$, the gap between states $\ket{\phi_k}$ and $\ket{\phi_{k+1}}$ with $k$ an odd number is $\beta$, but the gap between even and odd numbered states takes on a more complicated structure.

\section{Control Schemes}
In this section we first consider simple Gaussian and polynomial pulses, before using numerical control schemes.
We choose a low-dimension parameterization for the Gaussian and polynomial pulses and explore the fidelity landscape in this simplified setting, essentially a variational approach to optimizing the control scheme.
These low-parameter results demonstrate the complex time dynamics of the system, while numerical control schemes show that some limited improvement can be made using a higher-dimensional control manifold.

\subsection{Gaussian Pulse}
There are practical constraints on the implementation of physical control pulses, and simple smooth pulse shapes with reasonable bandwidths are preferred in experimental settings \cite{weidnerRobustQuantumControl2024}.
Motivated by these considerations, we first consider a simple control pulse with two control parameters $\left\{a,\omega \right\}$ and equation
\begin{align}\label{eq:pulse_gauss}
g_s(t) = a \exp \left[-
\frac{32(t-t_f / 2)^2}
{t_f^2}
\right]
\sin \left(
\frac{2 \pi t}{t_f}\omega
\right)^2.
\end{align}
In Fig. \ref{fig:fid_4_spins} we consider 4 spins and run 100 trials with $J_{i,i+1}\in [-1,1]$ and vary the pulse time $t_f \in {0.1,1.0,5.0}$, while performing a search using $a \in [-50,50]$ and $\omega \in [0.02,4]$.

Examples of the fidelity landscapes for the first trial (run number 1 in Fig. \ref{fig:fid_4_spins} a)) are shown in Fig. \ref{fig:fid_landscape_gauss_n4}. 
The fidelity landscape quickly becomes complicated as $t_f$ increases, with $t_f=5.0$ showing an extremely complicated landscape. For $t_f =0.1$ and $t_f =1.0$ the maximum fidelity is relatively independent of $\omega$, and high-fidelity bands are formed, showing that the amplitude of the pulse is critical to achieving high fidelity.

\begin{figure}[H]
\flushleft a) \\ \vspace{\figVspace pt}
\centering
\includegraphics{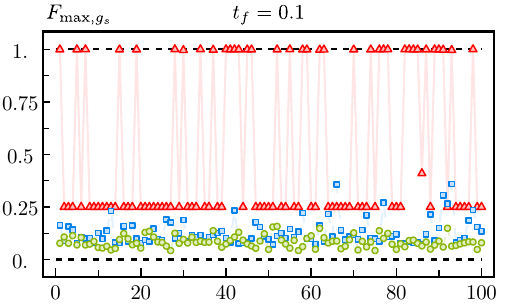}\\
\flushleft b) \\ \vspace{\figVspace pt}
\centering
\includegraphics{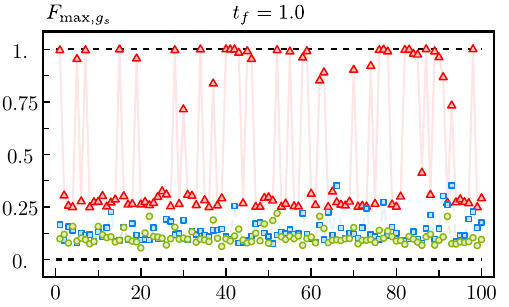}\\
\flushleft c) \\ \vspace{\figVspace pt}
\centering
\includegraphics{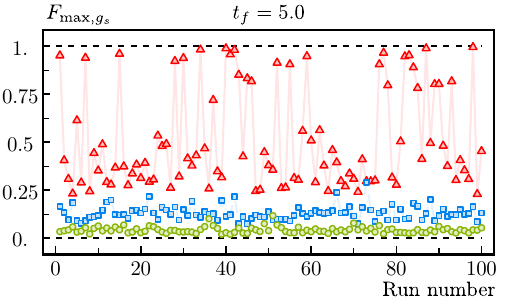}
\caption{Maximum fidelity of search using Gaussian pulse $F^{\text{max}}_{g_s}$ versus run number for 100 trials with $J_{i,i+1} \in {-1,1}$ and different spin numbers $N=4$ (red triangles), $N=8$ (blue squares) and $N=10$ (green diamonds).}
\label{fig:gauss_fid_n}
\end{figure}

\begin{figure}[H]
\flushleft a) \\ \vspace{\figVspace pt}
\centering
\includegraphics[width=\columnwidth]{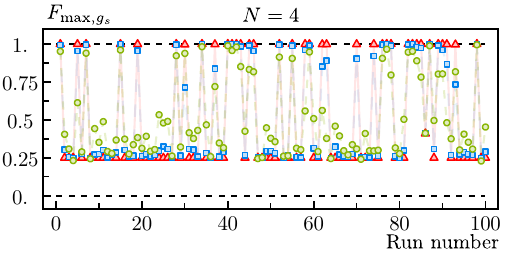}
\flushleft b) \\ \vspace{\figVspace pt}
\centering
\includegraphics[width=\columnwidth]{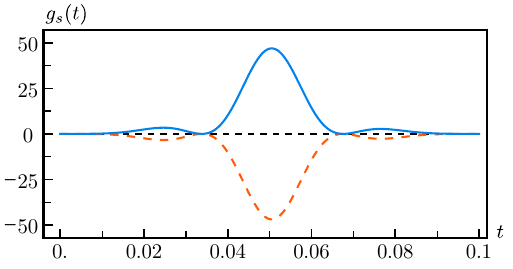}
\flushleft c) \\ \vspace{\figVspace pt}
\centering
\includegraphics[width=\columnwidth]{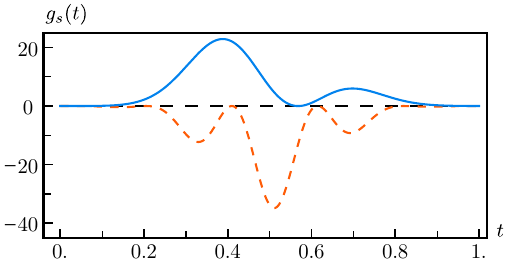}
\flushleft d) \\ \vspace{\figVspace pt}
\centering
\includegraphics[width=\columnwidth]{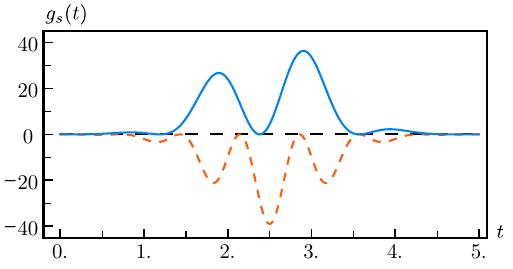}
\caption{a) Maximum fidelity achieved using a 2 parameter search over $a$ and $\omega$, with the Gaussian pulse from \Eqref{eq:pulse_gauss}. The system has $N=4$ spins and the search is run 100 times with different randomly chosen $J_{i,i+1}\in [-1,1]$ each time. The search is run for different final times $t_f$ with $t_f = 0.1$ (red circles), $t_f = 1.0$ (blue squares) and $t_f = 5.0$ (green triangles). b) - d) Examples of optimal pulses found using the search over $a$ and $\omega$ for 2 different runs (blue,dashed orange) and $t_f$. }
\label{fig:fid_4_spins}
\end{figure}

\begin{figure}[H]
\flushleft
\hspace{0.03\linewidth} a) \hspace{0.3\linewidth} b) \hspace{0.3\linewidth} c)\\ 
\centering
\includegraphics[width=168pt]{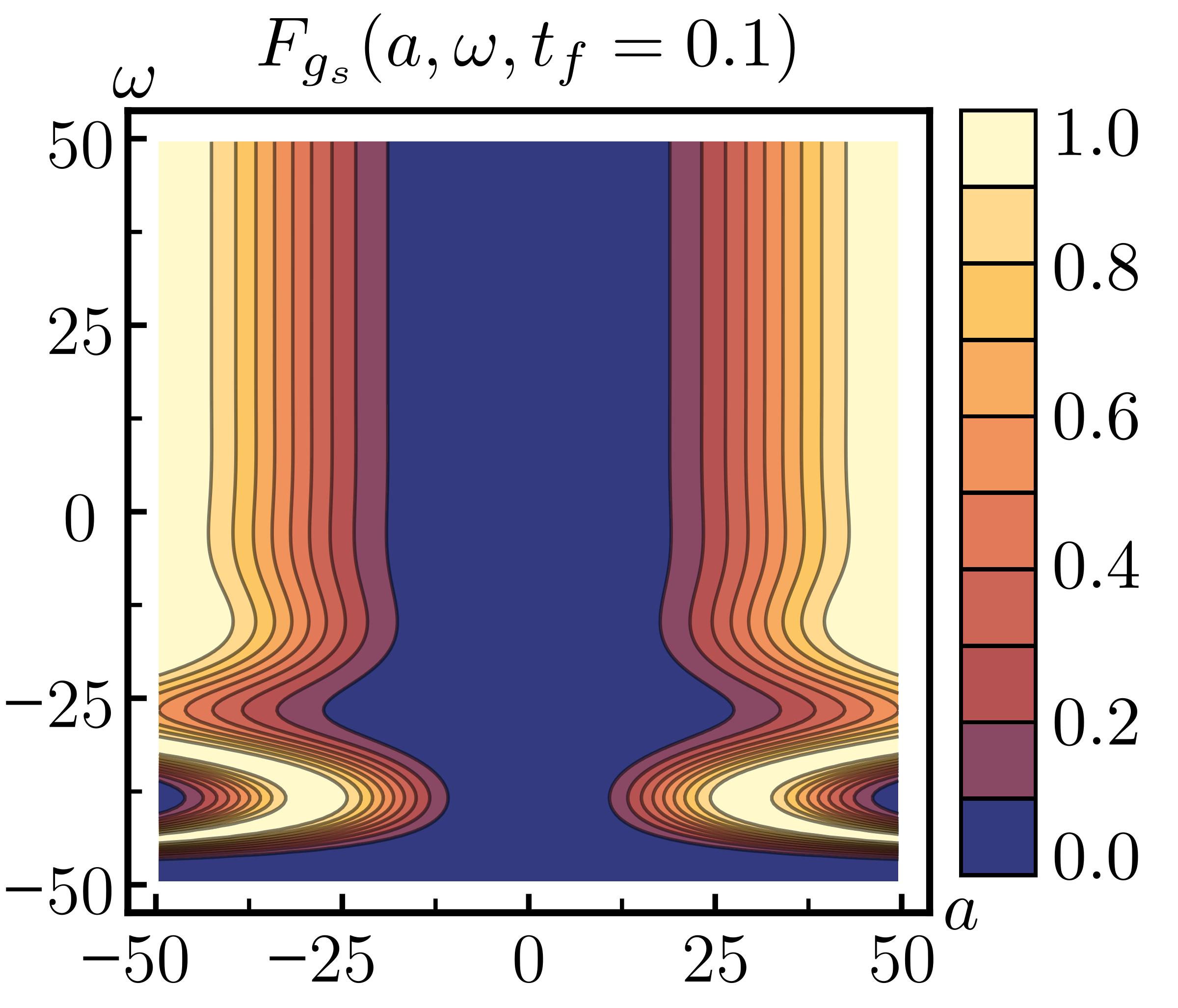}
\includegraphics[width=\imageSizeA\linewidth]{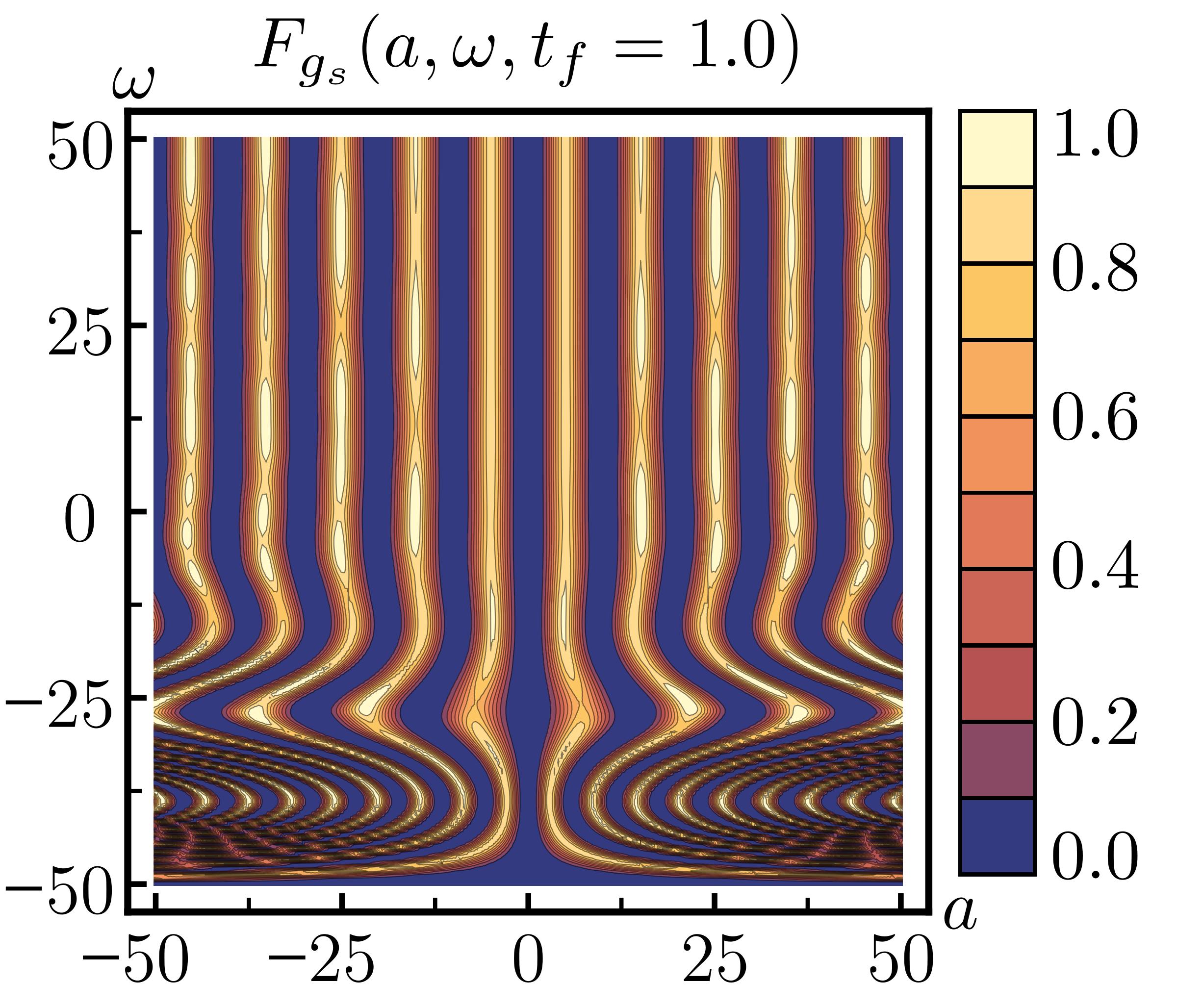}
\includegraphics[width=\imageSizeA\linewidth]{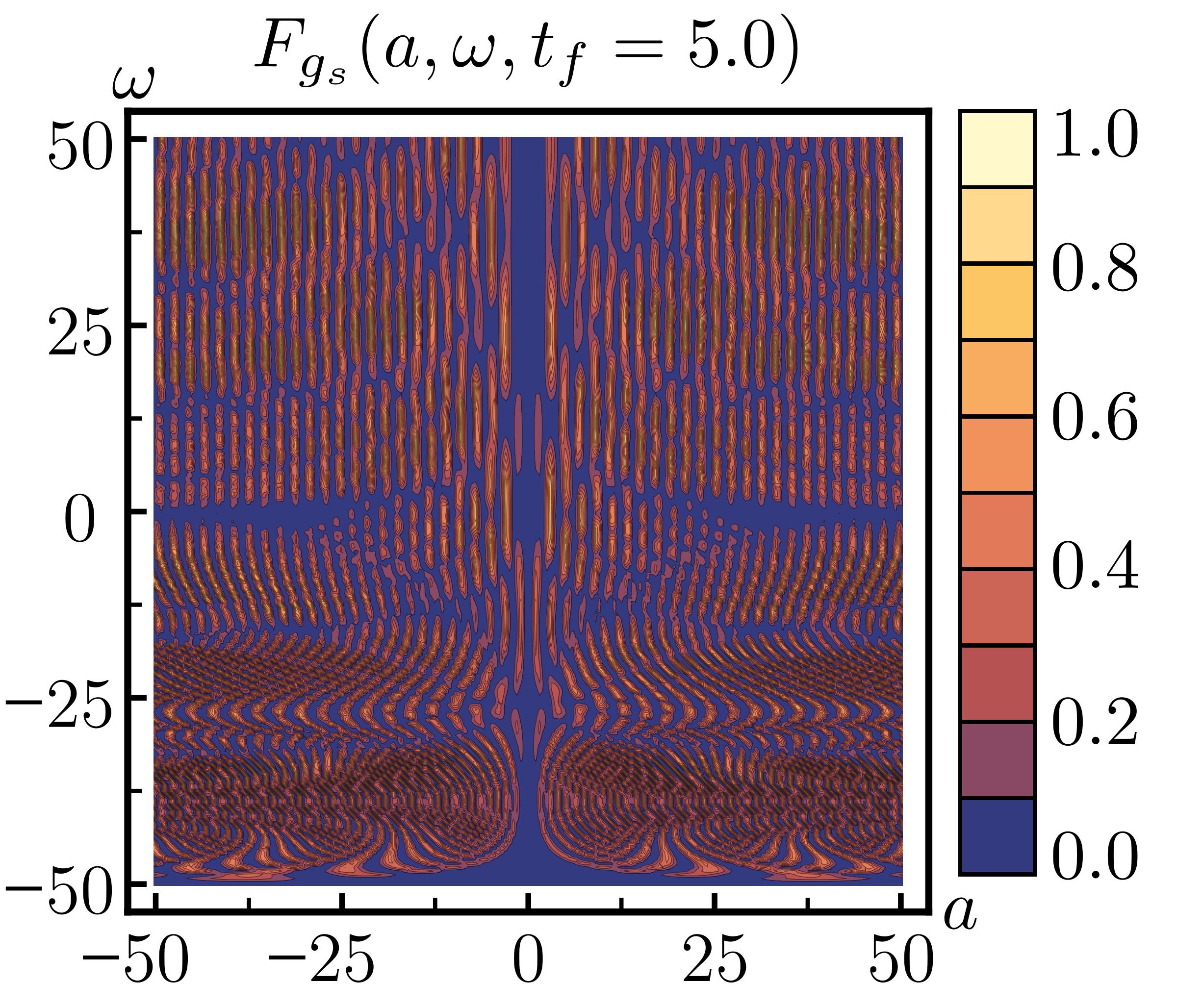}
\caption{Fidelity landscape using a 2 parameter search over $a$ and $\omega$, with the Gaussian pulse from \Eqref{eq:pulse_gauss}. The system has $N=4$ spins and we use different final times $t_f = 0.1$, $t_f = 1.0$ and $t_f = 5.0$. The couplings $J_{i,i+i}$ correspond with run number 1 in Fig. \ref{fig:fid_4_spins} a).}
\label{fig:fid_landscape_gauss_n4}
\end{figure}

In Fig. \ref{fig:gauss_fid_n} we consider different numbers of spins $n=4,8,10$ and perform a search on $a,\omega$ using the Gaussian pulse of \Eqref{eq:pulse_gauss}. 
For each of the 100 trials $J_{i,i+1}\in [-1,1]$, and again we vary the pulse time $t_f \in {0.1,1.0,5.0}$.
As the number of spins increases, the fidelity falls for each of the different $t_f$, with high fidelity for $N=4$ due to finite system effects.
In general, these results demonstrate that arbitrary excited state transfer in large spin systems is unlikely to be realized using a simple global Gaussian field on the transverse term.

\onecolumngrid

\twocolumngrid

\subsection{Polynomial Pulse}
In this section we consider a polynomial pulse and apply appropriate boundary conditions to produce a smooth pulse at initial and final times, while also allowing flexibility in choosing the number of free parameters for optimization.
We consider a polynomial pulse $g_p(t)$ given by
\begin{align}
g_{p}(t)=\sum_{j=1}^J a_j t^j
\end{align}
that satisfies the boundary conditions
$g_p(0) = \dot{g}_p(0) = \ddot{g}_p(0) = 0$,
$g_p(t_f) = \dot{g}_p(t_f) = \ddot{g}_p(t_f) = 0.$

We set $N_\lambda$ as the number of free parameters in the final pulse and choose a parameterization where the control parameters are fixed points along the control pulse, i.e. $\lambda_k = g_p[t = k\times t_f / (N_\lambda+1)]$ and $k=1\dots,N_\lambda$.
For example, if we have two control parameters $\{\lambda_1,\lambda_2\}$, we need a 7th order polynomial $(J=7)$ and obtain
\begin{align}\label{eq:pulse_poly}
g_{p}(t,\lambda_1,\lambda_2)=\frac{729 t^3 (t-t_f)^3}{8 t_f^7}
\left[\lambda_1 (3 t-2 t_f)+\lambda_2 (t_f-3 t)\right].
\end{align}
A plot of $g_{p}(t,-0.4,1.0)$ is shown in Fig. 4 and examples of the fidelity landscape using $N_\lambda=2$ components are shown in Fig. \ref{fig:fid_landscape_poly_n4}.
\begin{figure}[H]
\centering
\includegraphics[width=\linewidth]{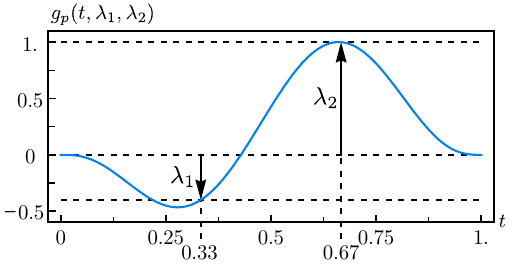}
\vspace{-15pt}
\caption{Fig. 2: Example of $g_{p}(t,\lambda_1,\lambda_2)$ with $\lambda_1=-0.4, \lambda_2=1.0, t_f=1.0$.}
\label{fig:plot_2}
\end{figure}
We first consider a brute force search for optimal $\lambda_1,\lambda_2$ for each trial run and in Fig. \ref{fig:poly_fid_4_spins} a) the maximum fidelity over each search is shown for $N_\lambda=4$ components $g_p$ with $\lambda_k \in [-30,30]$,for 100 trials of random interactions $J_{i,i+1} \in [-1,1]$.
We also considered a random search, whereby 1000 random guesses of $\vec\lambda$ with 2 and 4 components, and found little difference in performance to the brute-force search.
In the case of $t_f=5.0$, a difference in performance was found due to the complicated localized nature of the fidelity landscape as demonstrated in Fig. \ref{fig:fid_landscape_poly_n4} d).

\onecolumngrid

\begin{figure}[h]
 \flushleft \hspace{0.04\linewidth} a) \hspace{0.45\linewidth} b) \\ 
\centering \vspace{-5 pt}
\includegraphics[width=\imageSizeB\linewidth]{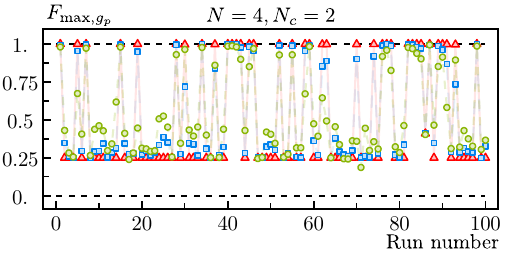}
\includegraphics[width=\imageSizeC\linewidth]{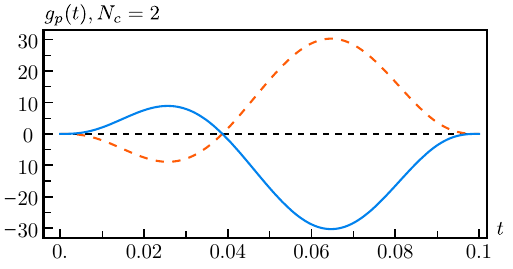}\\
\vspace{-5 pt}
\flushleft \hspace{0.04\linewidth} c) \hspace{0.45\linewidth} d) \\ 
\centering
\includegraphics[width=\imageSizeC\linewidth]{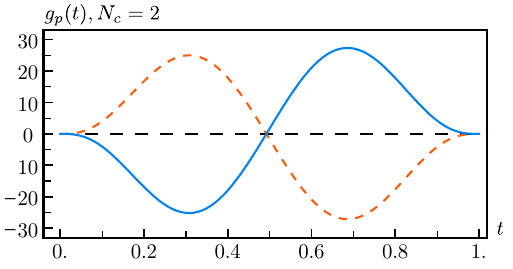}
\includegraphics[width=\imageSizeC\linewidth]{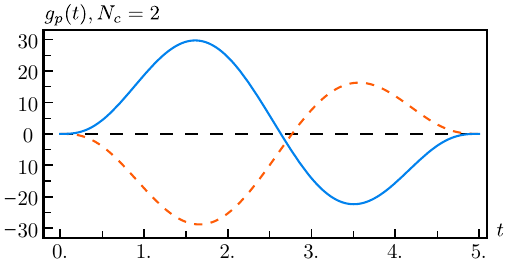}
\caption{a) Maximum fidelity achieved using a parameter search over $N_c=4$ components, $\lambda_1,\dots,\lambda_4$, with the polynomial pulse from \Eqref{eq:pulse_gauss}. The system has $N=4$ spins and the search is run 100 times with different randomly chosen $J_{i,i+1}\in [-1,1]$ each time. The search is run for different final times $t_f$ with $t_f = 0.1$ (red circles), $t_f = 1.0$ (blue squares) and $t_f = 5.0$ (green triangles). b) - d) Examples of optimal pulses found using the search over $\vec\lambda$ for 2 different interaction runs (blue,dashed orange) and $t_f$. }
\label{fig:poly_fid_4_spins}
\end{figure}

\begin{figure}[h]
\flushleft \hspace{0.025\linewidth} a) \hspace{0.3\linewidth} b) \hspace{0.3\linewidth} c)\\ 
\centering
\includegraphics[width=\imageSizeA\linewidth]{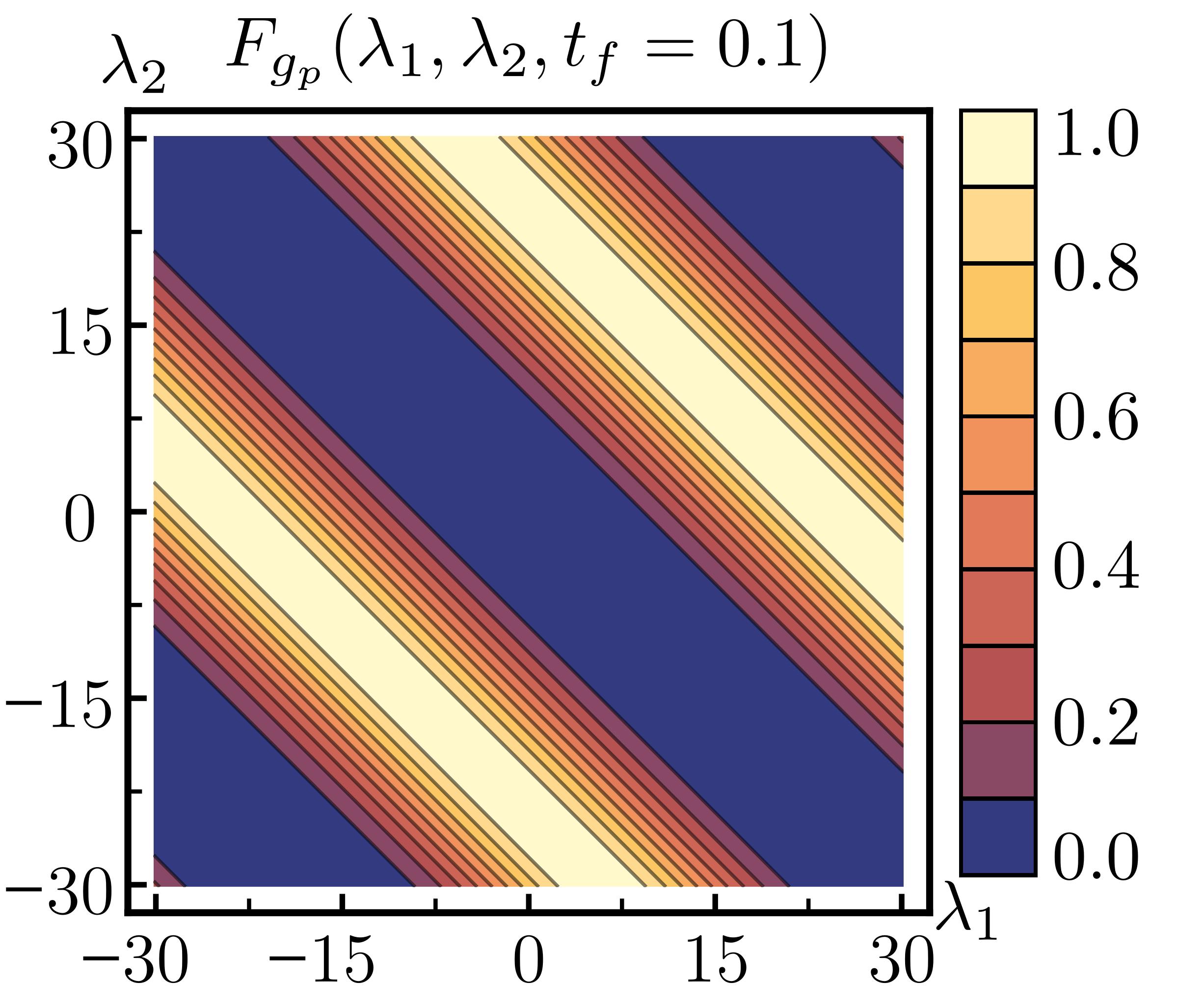}
\includegraphics[width=\imageSizeA\linewidth]{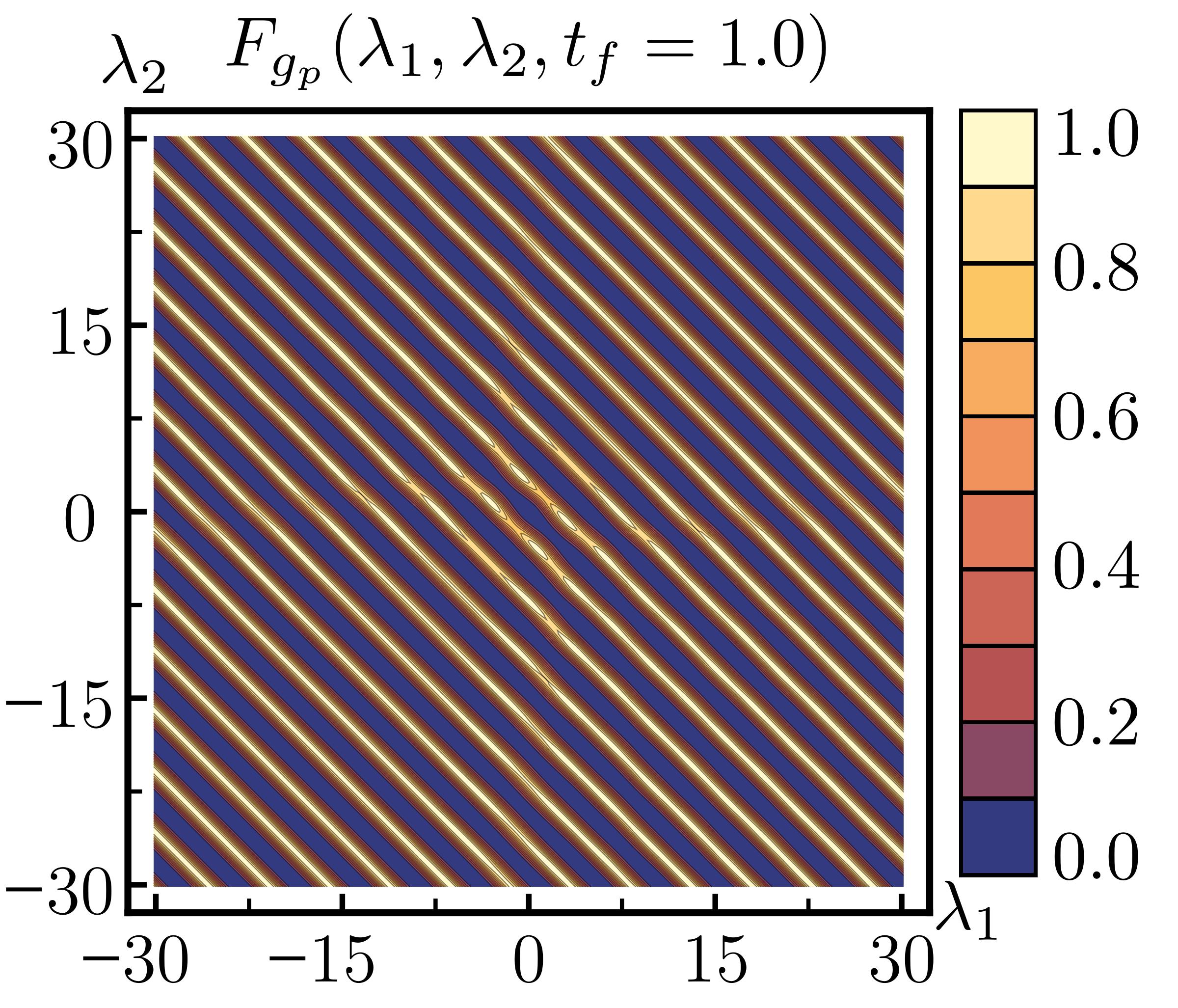}
\includegraphics[width=\imageSizeA\linewidth]{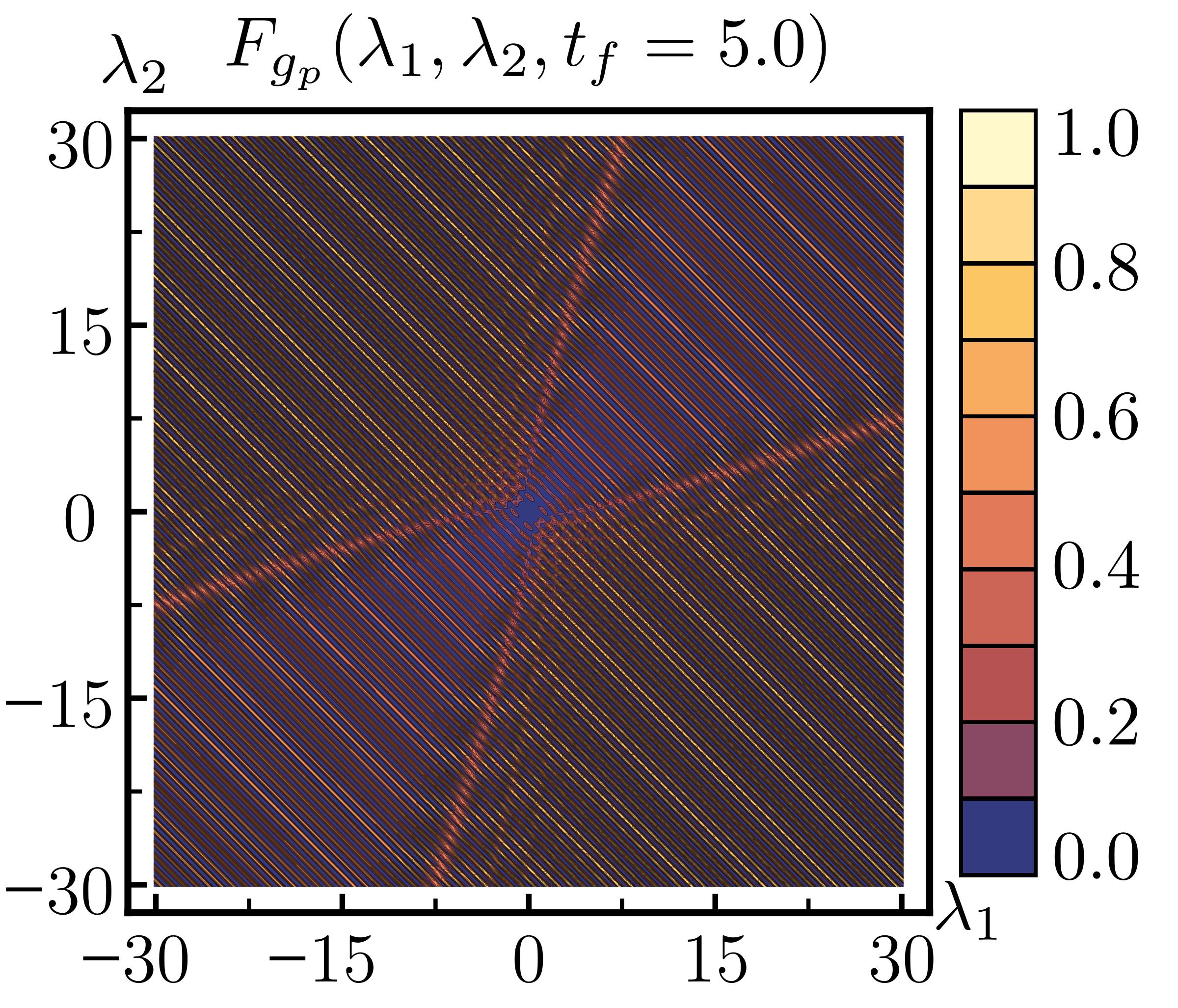}
\caption{ a) - c) Fidelity landscape using a 2 parameter search over $\lambda_1$ and $\lambda_2$, with the polynomial pulse from \Eqref{eq:pulse_poly}. The system has $N=4$ spins and we use different final times $t_f = 0.1$, $t_f = 1.0$ and $t_f = 5.0$. The couplings $J_{i,i+i}$ correspond with run number 1 in a).}
\label{fig:fid_landscape_poly_n4}
\end{figure}

\twocolumngrid

The fidelity landscapes of the Gaussian and polynomial pulses are similar as $t_f$ increases, where the landscape becomes very localized and sensitive to changes in the control parameters (see Fig. \ref{fig:fid_landscape_gauss_n4} and Fig. \ref{fig:fid_landscape_poly_n4})
For $t_f=0.1$ two distinct regions of high fidelity emerge in both pulse types, while for $t_f=1.0$ bands of high fidelity emerge.
The polynomial pulse shape has high-fidelity regions with a linear relationship between $\lambda_1$ and $\lambda_2$.
For example, in Fig. \ref{fig:fid_landscape_poly_n4} a) there is a strip of high fidelity around the line $\lambda_2+\lambda_1+25 \approx 0 $.
This demonstrates that the system is robust against changes in the control parameters for this region of parameter space.

The Gaussian and polynomial pulses have some similarity in their pulse shapes and their overall performance is very similar.
Later we will compare them directly to other purely numerical control pulses and show that they are close to optimal.

\section{Numerical Schemes}
We now consider purely numerical control schemes, the well known gradient-based GRAPE algorithm and the CRAB algorithm.
These state-of-the-art quantum control methods can offer fast and highly optimized control schemes, but can result in physically difficult to implement control pulses \cite{weidnerRobustQuantumControl2024,kochQuantumOptimalControl2022}.

\subsection{GRAPE}
In this section, we apply the GRAPE algorithm and find similar results to the previous control schemes using Gaussian and polynomial pulses \cite{khanejaOptimalControlCoupled2005,rossignoloQuOCSQuantumOptimal2023}.
The details of the GRAPE algorithm are discussed in \cite{khanejaOptimalControlCoupled2005}, but we briefly describe the method here.

The main idea is to discretize the control pulse in time and then write an approximation to the gradient for each step in time with respect to the fidelity (or any other measure of performance for the control task at hand).
Typically, the Hamiltonian is of the form
\begin{align}
\cH = \cH_0 + \sum_{k=1}^M u_{k}(t) \cH_k,
\end{align}
where $\cH_0$ is the time-independent free evolution of the system and $\cH_k$ are constant.
The time evolution is discretized into $N$ equal steps of duration $\Delta t$.
For the $j^{th}$ time step the Hamiltonian has $M$ adjustable parameters $u_{k,j}$ that we assume are constant over $\Delta t$, i.e. $u_{k,j}(t)=u_{k,j}$ for $t_{j-1}<t<t_j$ with $j=0,\dots,N$ and $k=0,\dots M$.
This allows us to write the time evolution as a simple matrix exponential, with the time evolution during the $j^{th}$ time step given by
\begin{align}
U_j = \exp \left[ -\frac{i}{\hbar} \Delta t \left(\cH_0 + \sum_{k=1}^M u_{k,j} \cH_k \right) \right].
\label{eq:bg_U_j}
\end{align}
The final state after the full time evolution is
\begin{align}
\ket{\Psi(t_f)} = \ket{\Psi_N} = U_N \dots U_1 \ket{\Psi_i}, 
\end{align}
and we define a state transfer cost function as
\begin{align}
C = 1 - \fabsq{ \braket{\Psi_f}{\Psi_N} }.
\end{align}
Now we calculate $\partial C / \partial u_{k,j}$ over the $j^{th}$ time interval to first order in $\Delta t$ \cite{khanejaOptimalControlCoupled2005}
\begin{align}
\frac{\partial C}{ \partial u_{k,j}}
&=
-2 i \Delta t \:
\text{Re} \Big[
\braket
{U_{j+1}^\dagger \dots U_N^\dagger\Psi_f}
{\:\cH_k U_j U_{j-1}\dots U_1 \Psi_0} \times
\nonumber \\
&\hspace{2cm} \braket{\Psi_N}{ \Psi_f}
\Big].
\label{eq:bg_grape_derv}
\end{align}
For closed quantum systems there is an efficient way to implement the gradient calculation.
Using current control parameters $u_{k,j}$, evolve $\ket{\psi_0} \rightarrow \ket{\psi_N}$. If the fidelity $\fabsq{\braket{\psi_T}{\psi_N}}>c$, we are done (where $c$ is a threshold fidelity we choose). Then evolve backwards $\ket{\psi_N}$ and $\ket{\psi_f}$ simultaneously using $U_N^\dagger$ a single $\Delta t$ step and calculate all $k$ derivatives using \Eqref{eq:bg_grape_derv}, to give the gradient for $j=1$.
Repeat the previous for all time steps, calculating the gradient for each $j$.
Update the control parameters $u_{k,j}\rightarrow u_{k,j} + \epsilon \frac{\partial C}{ \partial u_{k,j}}$, with a chosen small $\epsilon$ (in practice a simple line search can be used, or extended versions see \cite{numrecipes}). 
Repeat until the desired goal is achieved.

For the results considered here we discretize the control time with 10 and 100 steps to produce the GRAPE pulses.
Choosing more steps in general will achieve better results, but at the cost of increased complexity in the pulse shape \cite{khanejaOptimalControlCoupled2005}. 

\begin{figure}[H]
\flushleft a) \\ \vspace{\figVspace pt}
\centering
\includegraphics{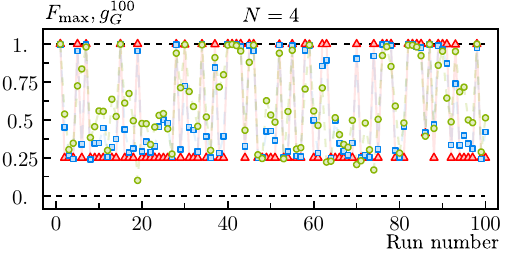}
\flushleft b) \\ \vspace{\figVspace pt}
\centering
\includegraphics{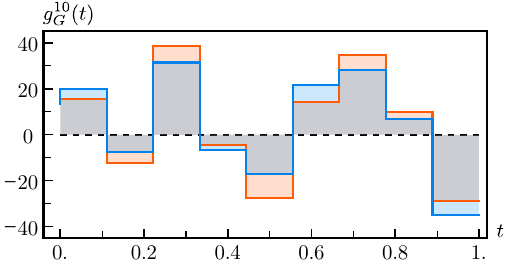}
\flushleft c) \\ \vspace{\figVspace pt}
\centering
\includegraphics{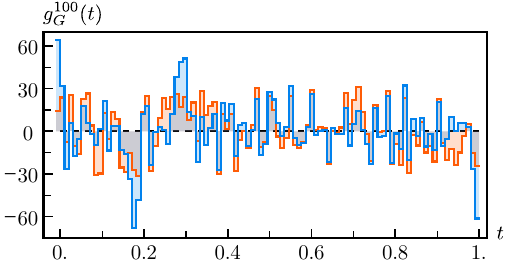}
\caption{a) Max fidelity found using 100 bin GRAPE with $N=4$ and different final times $t_f = 0.1$ (red circles), $t_f = 1.0$ (blue squares) and $t_f = 5.0$ (green triangles). 
b) \& c) Examples of the optimized GRAPE pulse with 10 and 100 bins for $t_f=1.0$. While the pulse shapes are quite different, the resulting difference in fidelity is negligible.}
\label{fig:grape_1}
\end{figure}

In Fig. \ref{fig:grape_1} a) the fidelity is shown for different run numbers and can be compared directly with Fig. \ref{fig:fid_4_spins} a) and Fig. \ref{fig:poly_fid_4_spins} a).
There is almost no difference in maximum fidelity achieved for $t_f=0.1$, and only some minor differences for $t_f=1.0$ and $t_f=5.0$.
For the longer control pulse times, the difference between the GRAPE pulse and the Gaussian and polynomial pulses is expected, since the fidelity landscape is increasing in complexity.
Interestingly, the simpler pulses overall perform similarly to the GRAPE pulses, suggesting fundamental limits on the controllability of the system.

\subsection{CRAB}
The CRAB algorithm was originally developed for many body systems, in particular systems that can be efficiently simulated using a time-dependent matrix renormalization group \cite{RevModPhys77259,doriaOptimalControlTechnique2011}.
The main idea of the algorithm is to make an ansatz for the control function that is constructed in a truncated basis, where we choose the elements of the truncated basis randomly, which allows the fidelity landscape to be explored in an efficient manner.
The basis functions can be any set that spans $L^2$, and we choose a finite number of them that span a subspace $S \subset L^2$.

For example, we can write our ansatz for the control function as
\begin{align}
g(t)=\sum_{i=1}^{N_c} c_i g_i(t)
\end{align}
where $S=\text{span} \{g_1(t),\dots,g_{N_c}(t)\} $.
An important point here is that $S$ should be of small dimension so that the optimization of the coefficients $c_1,\dots,c_{N_c}$ can be done fast.
If we fix this subset of basis functions $S$, then there is a risk that the algorithm may not explore the fidelity landscape effectively, and the algorithm could get caught in a local maximum (or minimum, depending on the control objective).
To overcome this limitation, a number of subsets $S_k$ are chosen with random choices for the basis functions $g_i^(k)$ in each $S_k$.
This leads to several control pulses $g^(k)(t) $ that are each built from their respective $S_k$, which are then optimized in parallel using standard search methods (e.g. Nelder-Mead simplex algorithm, Powell's method, Monte Carlo optimization, see \cite{mullerOneDecadeQuantum2022}).

A refinement to the CRAB algorithm is the dCRAB algorithm (dressed chopped random basis) that uses an iterative procedure to produce a new control pulse based on the optimal control pulse of the previous step and newly chosen random basis functions \cite{rachDressingChoppedrandombasisOptimization2015}.
In dCRAB, we choose a set of basis functions $S_k$ and produce an optimized pulse as discussed previously, then update the pulse with new basis states, i.e. the initial optimized pulse is now dressed with new basis states.
For the $jth$ step in the dCRAB procedure, with $g^{j-1}(t)$ be the initial optimized  control pulse, we update it to
\begin{align}
g^j(t)=g^{j-1}(t)+\sum_{i=1}^{N_{c}} c^j_i f^j_i(t)
\end{align},
where $f^j_i(t)$ are new randomly chosen basis functions.
We then optimize the new pulse (i.e. the coefficients $ c^j_i$) and repeat the procedure until we have convergence to our control goal.
This allows dCRAB to avoid local false traps in the fidelity landscape, where these false traps are due to the use of a truncated basis. 
dCRAB has been shown to avoid these false traps by adding a new random basis function to the pulse caught in a false trap \cite{rachDressingChoppedrandombasisOptimization2015}.
The effect of different choices of basis functions has been explored in \cite{Roleofbasesinquantumoptimalcontrol}, and for the results presented here we use the Fourier basis.

\begin{figure}[H]
\flushleft a) \\ \vspace{\figVspace pt}
\centering
\includegraphics{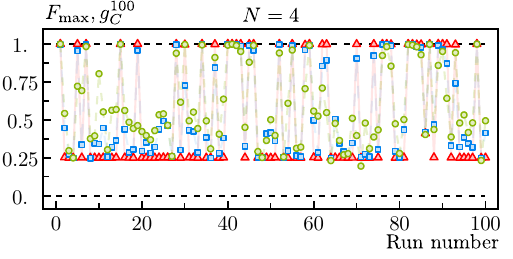}
\flushleft b) \\ \vspace{\figVspace pt}
\centering
\includegraphics{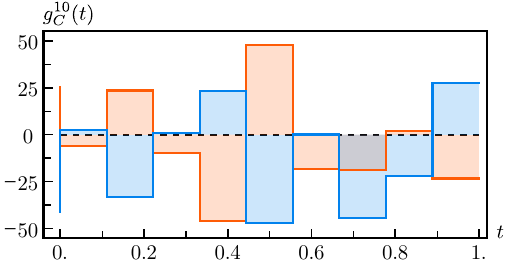}
\flushleft c) \\ \vspace{\figVspace pt}
\centering
\includegraphics{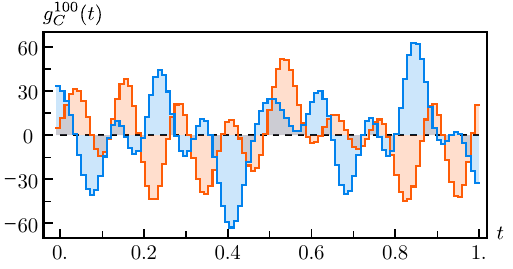}
\caption{a) Max fidelity found using 100 bin dCRAB with $N=4$ and different final times $t_f = 0.1$ (red circles), $t_f = 1.0$ (blue squares) and $t_f = 5.0$ (green triangles). 
b) \& c) Examples of the optimized dCRAB pulse with 10 and 100 bins for $t_f=1.0$.}
\label{fig:crab_1}
\end{figure}

The results of applying dCRAB to our control problem are shown in Fig. \ref{fig:crab_1}.
Later we compare all the control schemes used (Fig. \ref{fig:conclusion}), and find that for $t_f=1.0,5.0$ dCRAB has some advantage over the other schemes.
This is likely due to subspace search method it employs, but note that all the control schemes have poor fidelity for $t_f=5.0$.

\section{Conclusion}
\begin{figure}[H]
\flushleft
a)\\ \vspace{-15 pt}
\includegraphics{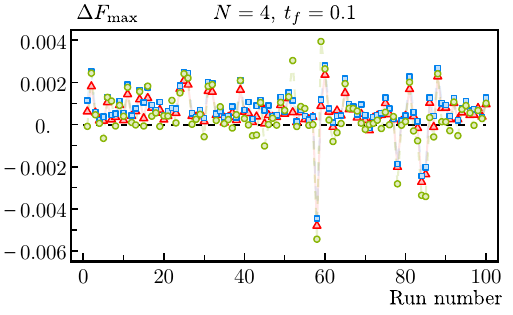}
\flushleft
b)\\ \vspace{-20  pt}
\includegraphics{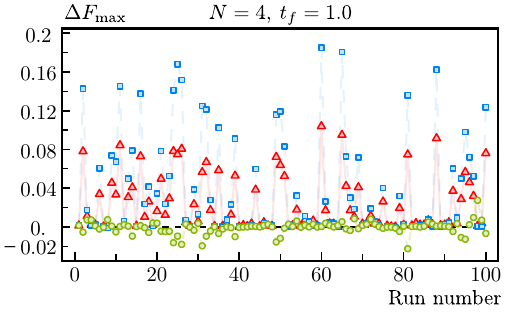}
\flushleft
c)\\ \vspace{-20 pt}
\includegraphics{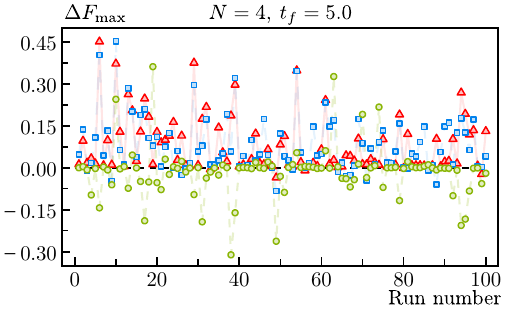}
\vspace{-12 pt}
\caption{Comparison of the difference in fidelity for 4 spins using the dCRAB pulse with 100 bins ($g_C^{100}$) as a reference; difference between the reference and polynomial pulse with 4 components $\Delta F_{\text{max}}(g_C^{100},g_p^{4}) $ (red triangles), difference between the reference and Gaussian pulse $\Delta F_{\text{max}}(g_C^{100},g_s) $ (blue squares) and difference between the reference and GRAPE pulse with 100 steps $\Delta F_{\text{max}}(g_C^{100},g_G^{100}) $ (green circles).
a) $t_f=0.1$, b) $t_f=1.0$ and c) $t_f=5.0$.}
\label{fig:conclusion}
\end{figure}

We have compared several different quantum control techniques for subspace-to-subspace transfer and found that the random interactions and pulse time play a critical role in allowable dynamics.
In Fig. \ref{fig:conclusion} the difference in fidelity between the various approaches is shown, and in general the different techniques agree with each other for short timescales but the topology of each optimal pulse can be very different.
For longer timescales, dCRAB can have some advantages due to how it explores the fidelity landscape, but as illustrated in Fig. \ref{fig:fid_landscape_n8} the fidelity landscape is in general highly complex making this control task difficult.
This is also why gradient methods would not be expected to perform well in this setting.

There are two main problems when trying to achieve state transfer in this system, the transverse term acts globally and the random nearest neighbor couplings dramatically change the time evolution dynamics.
For short pulse times and only four spins, the maximum fidelity achieved is generally 0.25 or 1, regardless of the control scheme employed.
The effects of the finite system size are strongest here, and as shown in Fig. \ref{fig:gauss_fid_n}, as the number of spins increases the maximum fidelity achieved falls.

In Appendix A Fig. \ref{fig:fid_landscape_n8} the fidelity landscapes for the Gaussian and polynomial pulses are shown for $N=8$ spins, and examples of the optimal pulses.
In the results for both pulses the fidelity landscape is increasing in complexity as $t_f$ is increased, and the infidelities achieved for $N=8$ are much worse than for $N=4$, agreeing with Fig. \ref{fig:gauss_fid_n}.
We speculate that setting the number of spins to be large, or at least $N>10$, we would not be able to engineer the subspace transfer task.
Machine learning approaches have had success in quantum annealing in similar systems \cite{hibat-allahVariationalNeuralAnnealing2021,berezutskiiProbingCriticalityQuantum2020}, but in the control task attempted here, it is unlikely that they could provide a solution without further modification to the Hamiltonian.

Our results suggest that time-dependent control over the couplings between the nearest-neighbor terms is necessary to engineer a solution for a given $t_f$.
One caveat here is that for very long $t_f$, and with a very large number of time steps, the average fidelity across all runs can be close to 1.
For example, $t_f=10^4$ with $10^3$ steps using the GRAPE algorithm does achieve close to 1 for almost all runs, but the resulting pulse is extremely complicated and not useful in a practical setting.
This control task considered here is very different from control problems involving the ground state, where adiabatic or other methods can be used \cite{guery-odelinShortcutsAdiabaticityConcepts2019,mitraQuantumQuenchDynamics2018,rajakQuantumAnnealingOverview2022a, espinosInvariantbasedControlQuantum2023,stefanescuRobustImplicitQuantum2024,orozco-ruizQuantumControlQuantum2024a}.
To achieve the kind of subspace transfer attempted here, one would need to consider time-dependent control over the coupling strengths in either the nearest-neighbor term, the transverse term, or both.
Another strategy would be to consider further time-dependent control terms in the Hamiltonian, for example next nearest-neighbor or terms that couple the spins further.

Another possible direction for future work would be to consider the control of this system through the dynamical Lie algebra \cite{dalessandroIntroductionQuantumControl2007a,2024wiersemaClassificationDynamicalLie}.
Recent work has shown that the transverse Ising model with constant coupling strengths is subspace controllable, meaning that while arbitrary state transfers are not possible, there are certain state transfers that are allowed \cite{2025dalessandroControllabilityPeriodicQuantum}.
It would be interesting to apply the techniques developed in that work to the model considered in this paper.
\begin{acknowledgments}
We wish to acknowledge support from IKUR PCI2022-132984, EPIQUS GA 899368 (H2020-FETOPEN-2018-2020), PID2021-126273NB-I00 project of MCIN/AEI/10.13039/501100011033 and ERDF “A way of making Europe”, the Basque Government through Grant No. IT1470-22, and UE and GCP credits. 
\end{acknowledgments}

\appendix
\onecolumngrid

\section{Fidelity landscapes for Gaussian and Polynomial pulses with $N=8$}
\begin{figure}[H]
\flushleft
\hspace{0.03\linewidth} a) \hspace{0.3\linewidth} b) \hspace{0.3\linewidth} c)\\ 
\centering
\includegraphics[width=\imageSizeA\linewidth]{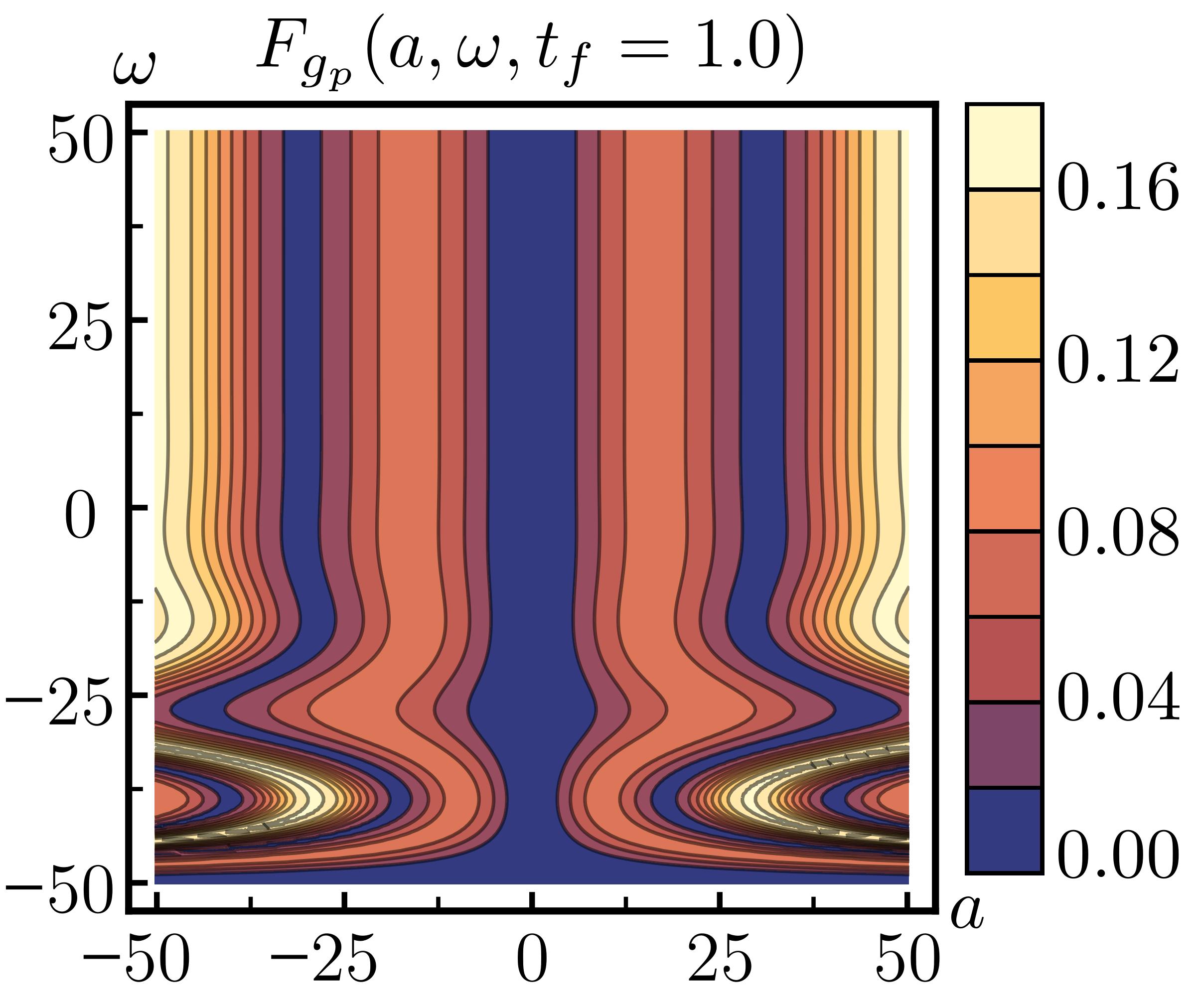}
\includegraphics[width=\imageSizeA\linewidth]{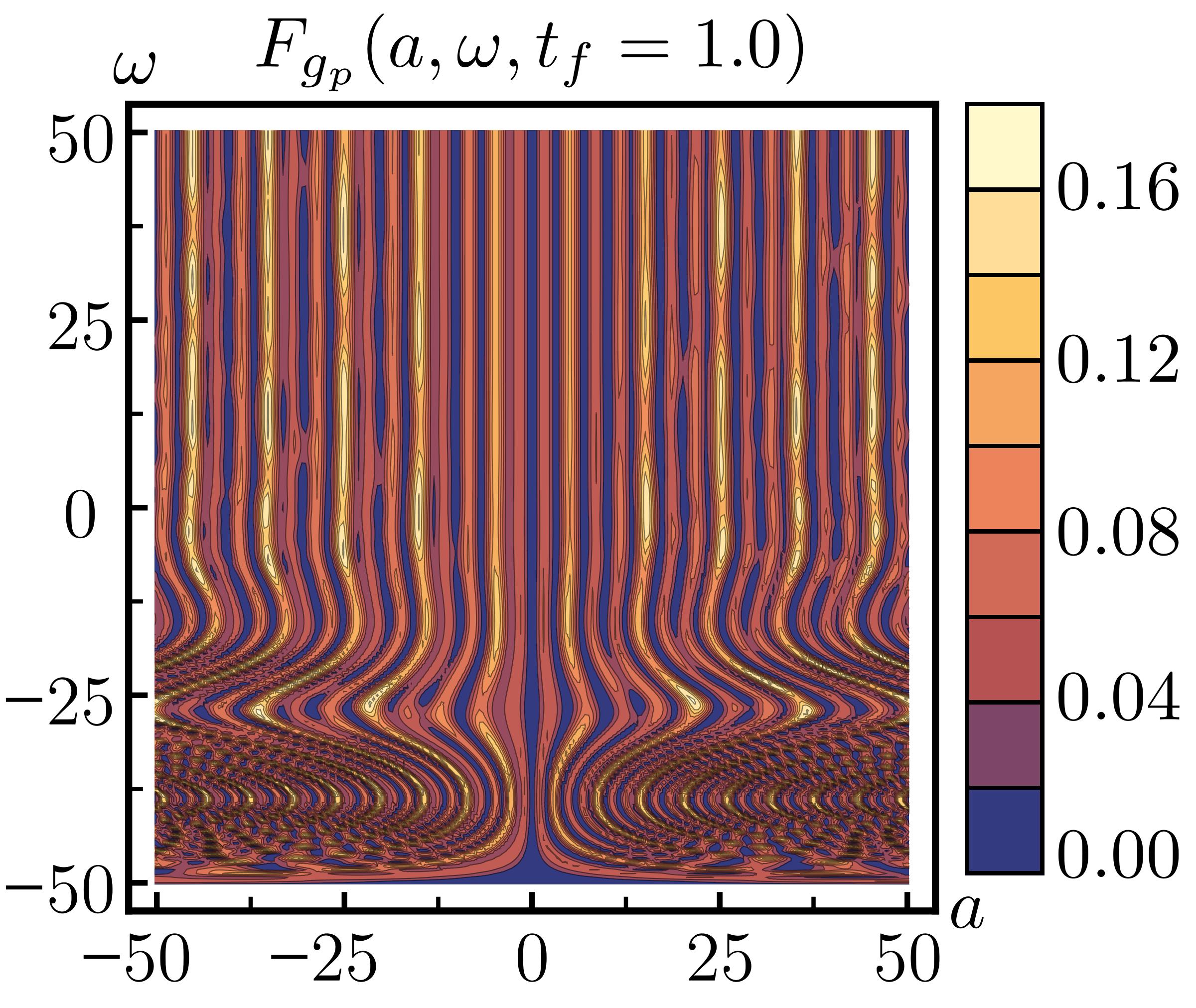}
\includegraphics[width=\imageSizeA\linewidth]{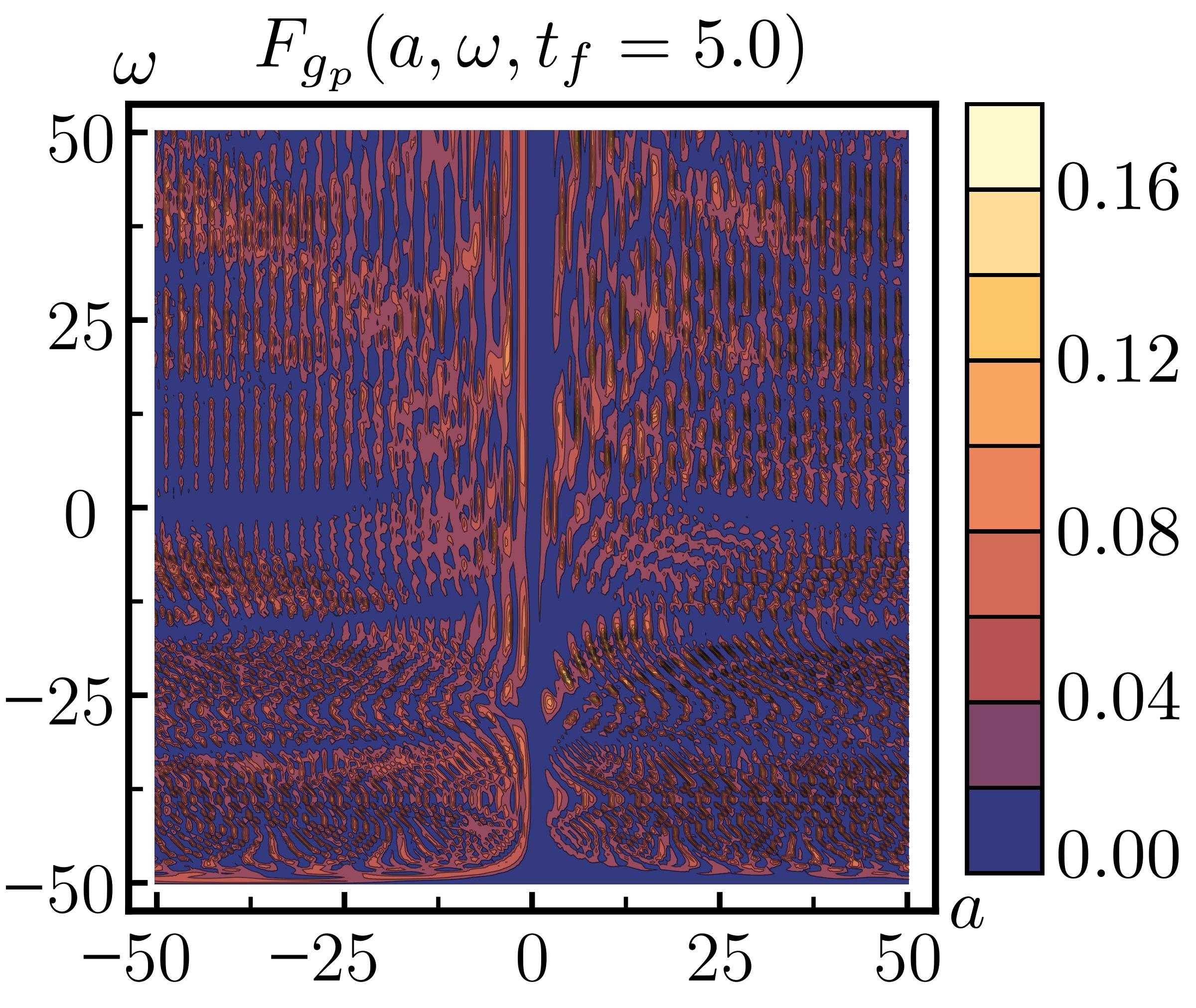}\\
\flushleft
\hspace{0.03\linewidth} d) \hspace{0.3\linewidth} e) \hspace{0.3\linewidth} f)\\
\centering
\includegraphics[width=\imageSizeA\linewidth]{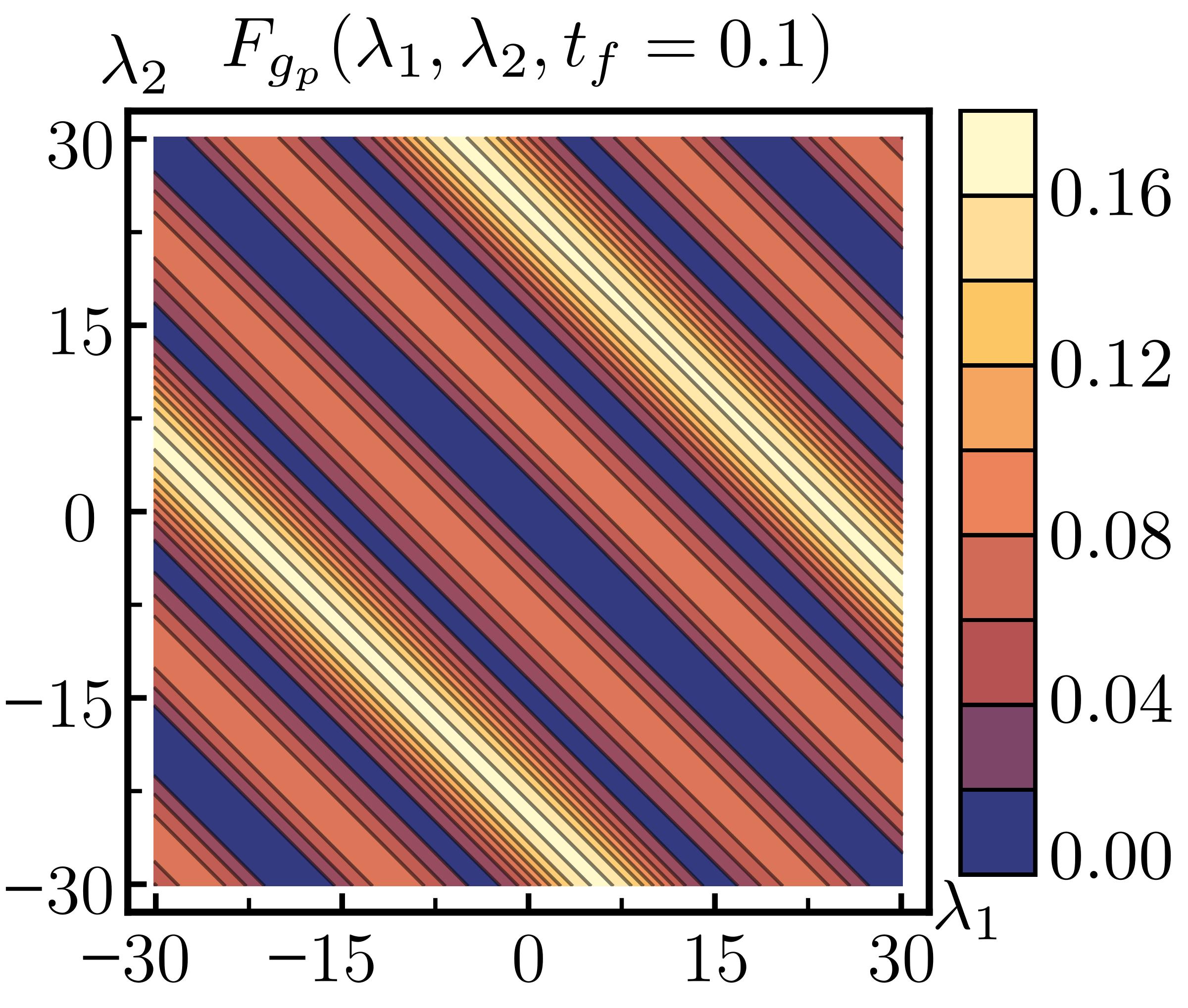}
\includegraphics[width=\imageSizeA\linewidth]{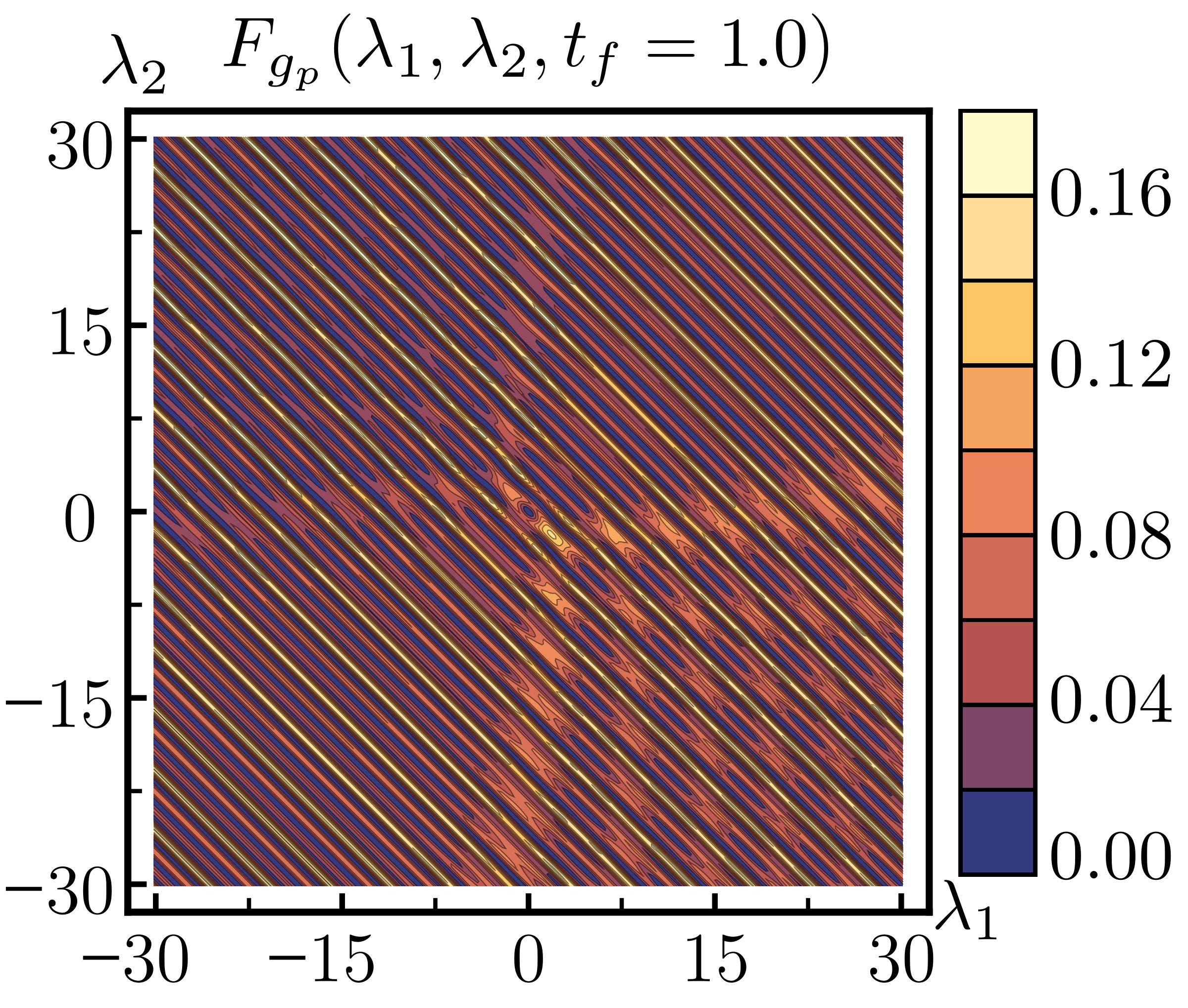}
\includegraphics[width=\imageSizeA\linewidth]{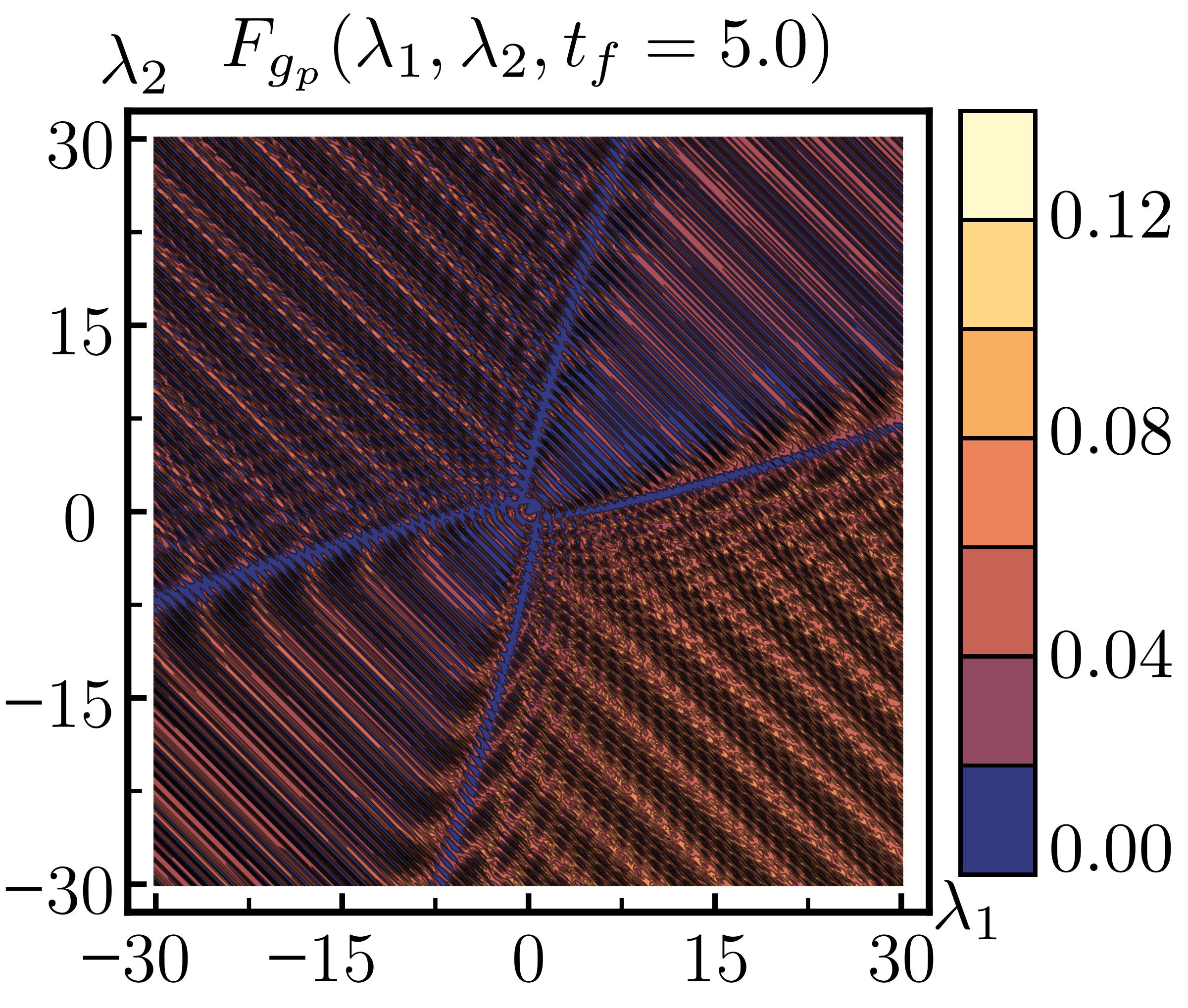}\\
\flushleft \hspace{0.0\linewidth} g) \hspace{0.475\linewidth} h) \\ 
\vspace{-10 pt}
\centering
\includegraphics[width=0.49\linewidth]{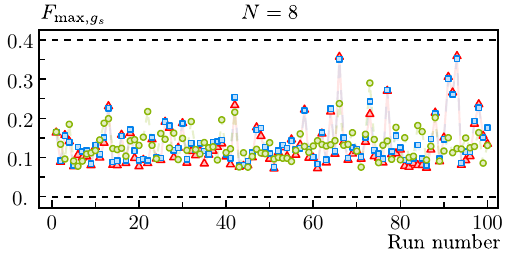}
\includegraphics[width=0.49\linewidth]{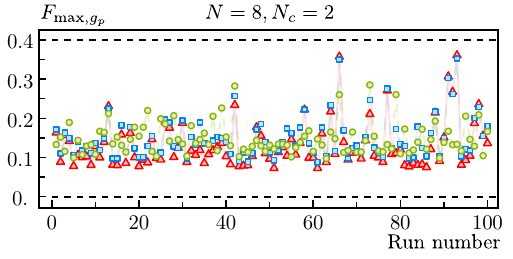}
\\
\flushleft \hspace{0.0\linewidth} i) \hspace{0.4825\linewidth} j) \\ 
\vspace{-10 pt}
\centering
\includegraphics[width=0.49\linewidth]{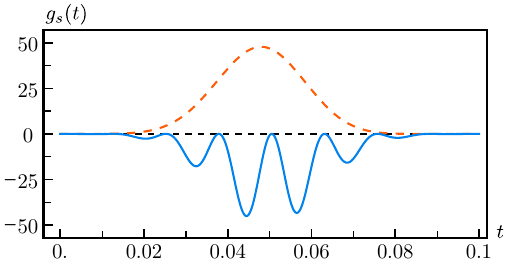}
\includegraphics[width=0.49\linewidth]{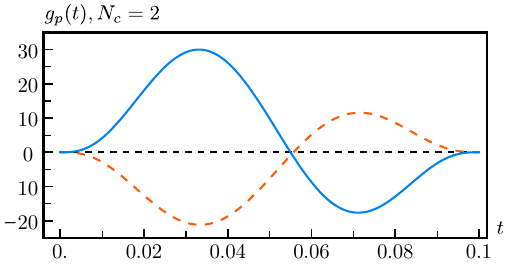}
\caption{a) to c) Comparison of the fidelity landscape using the Gaussian pulse (\Eqref{eq:pulse_gauss}) with $N=8$ spins) Different final times shown in columns $t_f = 0.1,1.0,5.0$. $J_{i,i+i}$ correspond with run number 1 in g) and h).
d) to f) Comparison of the fidelity landscape for 2 components polynomial pulse (\Eqref{eq:pulse_poly}) with $N=8$ spins and different $t_f$.
g)/h) Maximum fidelity using a parameter search over 2 components the Gaussian pulse ( \Eqref{eq:pulse_gauss}) / polynomial (\Eqref{eq:pulse_poly}) for $N=8$ spins and different $t_f$ with $t_f = 0.1$ (red circles), $t_f = 1.0$ (blue squares) and $t_f = 5.0$ (green triangles). 
i) and j) Example of 2 optimal Gaussian/polynomial pulses for $t_f=0.1$ and 2 different runs (blue,dashed orange).}
\label{fig:fid_landscape_n8}
\end{figure}

\twocolumngrid

\nocite{*}

\bibliography{paper}

\end{document}